\documentstyle[aps,prl,preprint,tighten,epsfig]{revtex}
\begin{document}
\title{A possible new phase of commensurate insulators with disorder:
the Mott Glass}
\author{E. Orignac$^1$, T. Giamarchi$^2$ and P. Le Doussal$^3$}
\address{$^1$ Physics Department, Rutgers
University, Piscataway, NJ 08855, U.S.A.}
\address{$^2$
Laboratoire de physique des Solides, CNRS UMR 8502,
UPS B{\^a}t 510, 91405 Orsay, France}
\address{$^3$ CNRS-Laboratoire de Physique Th{\'e}orique de L'Ecole
Normale Sup{\'e}rieure, 24, Rue Lhomond, 75231 Paris, France}
\date{\today}
\maketitle

\begin{abstract}
A new thermodynamic phase resulting from the
competition between a commensurate potential and disorder in
interacting fermionic or bosonic systems is predicted. It requires
interactions of finite extent. This phase,
intermediate between the Mott insulator and the Anderson insulator, is
both incompressible and has no gap in the conductivity. The
corresponding phase is also predicted for commensurate classical
elastic systems in presence of correlated disorder.
\end{abstract}

\narrowtext

The interplay between disorder and interactions gives rise to
fascinating problems in condensed matter physics. Although
the physics of disordered noninteracting systems is by now well
understood, when interactions are included the problem is largely
open, with solutions existing for one dimensional systems
\cite{giamarchi_loc_ref,feigleman_disorder_commensurate}
or through approximate scaling
\cite{finkelstein_localization_interactions,belitz_localization_review,%
fisher_bosons_scaling}
or mean-field methods
\cite{dobrosavljevic_infinited_disorder}. A particularly
interesting situation occurs when the non-disordered system
possesses a gap. This arises in a large number of systems
such as disordered Mott insulators
\cite{shankar_spinless_conductivite,fujimoto_mott+disorder_1ch,mori_scba,%
paalanen_silicon_phosphorus},
systems with external (Peierls or spin-Peierls
systems \cite{grenier_cugesio3}) or internal commensurate potential
(ladders or spin ladders
\cite{orignac_ladder_disorder,azuma_zinc_doping,carter_lacasrcuo,%
fujimoto_mott+disorder_2ch}, disordered spin 1 chains
\cite{monthus_spin1_ranexchange,kawakami_dopeds=1}).
This situation is also relevant for classical
problems such as elastic systems subjected to both periodic
potential and correlated disorder, as encountered e.g. in vortex
lattices in superconductors \cite{blatter_vortex_review}.

Although in some cases an infinitesimal disorder can suppress the gap
\cite{shankar_spinless_conductivite} due to Imry-Ma effects, in most
cases, a finite amount of disorder is needed to induce gap closure. In
the latter case, the complete description of the gap closure and of the
physics of the resulting phases is extremely difficult with the usual
analytic techniques such as perturbative renormalization group (RG),
due to the absence of a weak coupling fixed point. In $d=1$, where
attempts in solving this problem could be made, it was believed
\cite{fujimoto_mott+disorder_1ch,mori_scba} that two phases
existed: a weak disorder phase where the gap is robust and the system
has all the characteristics of the pure gaped insulator, and a strong
disorder phase where the gap is totally washed out by disorder and the
system is a simple compressible Anderson insulator. However the
techniques used so far are either restricted to special points
\cite{mori_scba} or suffer from serious limitation: the simple
perturbative RG of \cite{fujimoto_mott+disorder_1ch} has to be used
outside its regime of validity to describe the transition between two
strong coupling phases. Furthermore up to now mostly onsite
interactions have been studied.

\begin{figure}
 \centerline{\epsfig{file=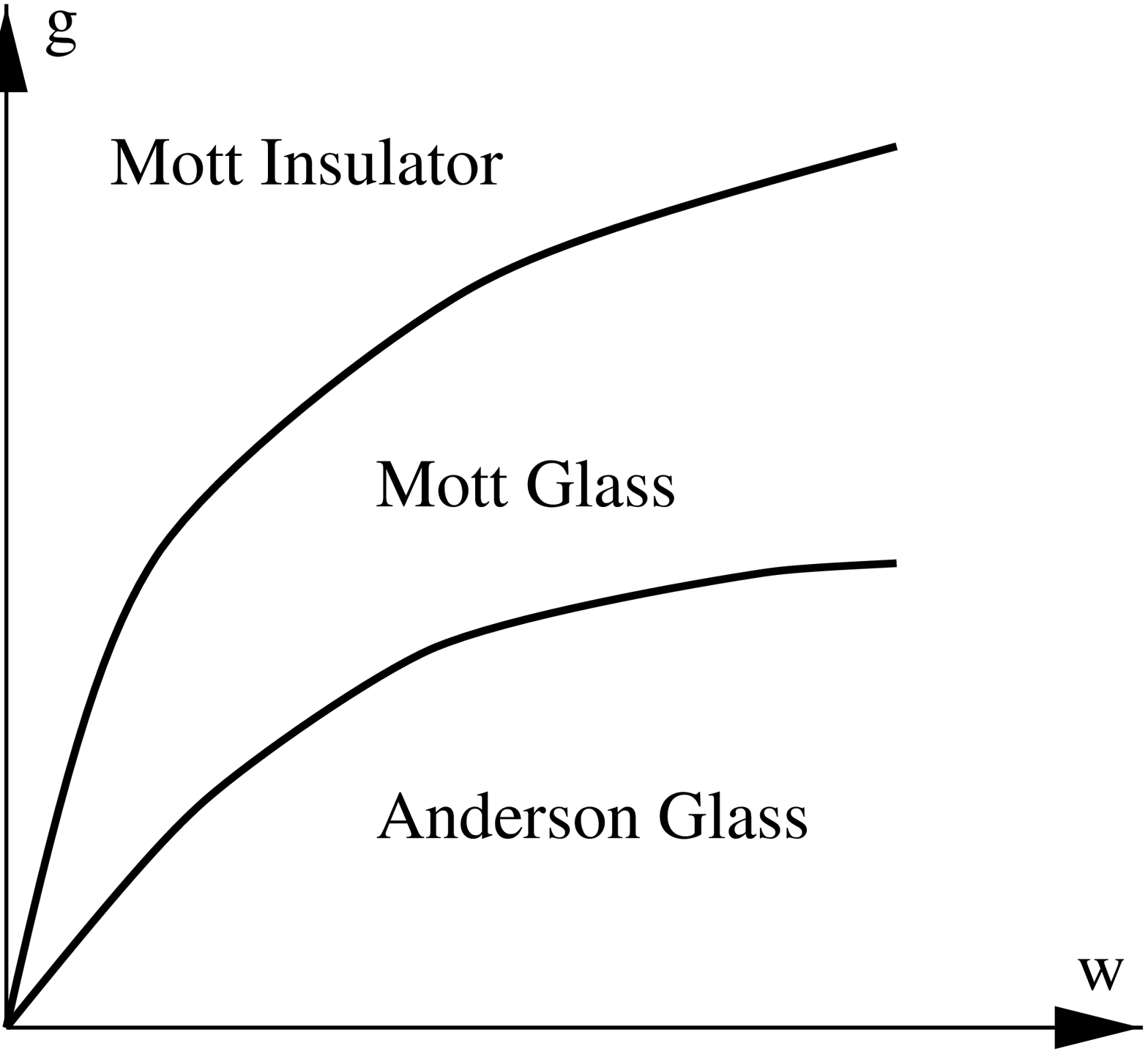,angle=-90,width=6cm}}
  \caption{\label{fig:phases} The three phases as defined in the text
  in the disorder $W$, commensurate potential $g$ plane for a
  fixed value of the interactions. Both MI and MG are incompressible.
  Both MG and AG have no gap in the conductivity.}
\end{figure}

In this Letter we thus reexamine this problem using
better suited methods that capture some nonperturbative effects: a variational
calculation and a functional renormalization group method.
We focus first, for simplicity, on
one dimensional interacting spinless fermions in the presence of a
commensurate potential \cite{periodic_mott}.
As argued below we expect similar physics to hold for systems with
spins. Our main finding is that, in addition to the
above mentioned phases, an intermediate phase exists, as shown in
Fig.~\ref{fig:phases}.

Quite remarkably although this phase possesses some of
the characteristics of an Anderson insulator, namely a non-zero ac
conductivity at low frequency, it remains \emph{incompressible}.
Since it is also dominated by disorder, we call it a Mott glass. We
discuss here the physical characteristics of this novel phase and
its interpretation for fermionic as well as bosonic systems.
We argue that such a phase is not
specific to $d=1$ and discuss its generalization to higher
dimension both for quantum and classical systems.

Disordered interacting spinless fermions submitted to a commensurate periodic
potential are described in one dimension by
\widetext
\begin{displaymath}
H= \int dx \left[-\imath \hbar v_F (\psi_+^\dagger
  \partial_x\psi_+-\psi_-^\dagger \partial_x\psi_-)  - g
  (\psi^\dagger_+\psi_-+\psi^\dagger_-\psi_+) + \mu  (x) \rho(x) \right]
  +\int dx_1 dx_2 V(x_1-x_2) \rho(x_1)\rho(x_2)
\end{displaymath}
\narrowtext
where $\pm    $ denote fermions with momentum close to $\pm    k_F$, as
is standard in $d=1$. $V$ is the interaction, $\mu    (x)$ an
on site random potential and $g$ the strength of the periodic
potential opening the gap. In $d=1$ it is convenient to use
a boson representation of the
fermion operators \cite{haldane_bosonisation}.
This leads to the action
\widetext
\begin{equation}\label{bosonized-hamiltonian}
S/\hbar = \int dx d\tau \left[ \frac1{2\pi K}\left[
\frac1{v}(\partial_\tau \phi)^2 + v (\partial_x \phi)^2
\right]  - \frac{g}{\pi\alpha\hbar} \cos 2\phi
+ \frac{\xi(x)}{2\pi\alpha\hbar}  e^{i 2 \phi(x)} + \text{c.c.}\right]
\end{equation}
\narrowtext
where $\alpha$ is a short distance cutoff (of the order of a lattice
spacing) and the field $\phi$ is related to the total density of
fermions by $\rho(x)  = -\nabla \phi/\pi$. The interactions are
totally absorbed in the Luttinger liquid coefficients $v$ (the
renormalized Fermi velocity) and $K$, a number that controls the
decay of the correlation functions. The noninteracting case
($V=0$) gives $K=1$ and $v=v_F$. We focus in the following on
the repulsive ($V >0$) case for which $0 < K < 1$ depending on the
strength and range of $V$. For weak disorder one can separate in
the random potential the Fourier components close to $q\sim 0$
(forward scattering) and $q\sim 2k_F$ (backward scattering). We
have retained here, for simplicity, only the backward scattering.
Forward scattering, studied in \cite{orignac_commensurate_long},
does not lead to qualitative changes in the physics presented
here. The backward scattering is
modeled by the complex gaussian random variable $\xi$ such that
$\overline{\xi(x)\xi^*(x')}= W \delta(x-x')$.

Perturbatively both the disorder (in the absence of commensurate
potential) and the commensurate potential (in the absence of disorder)
are relevant (respectively for $K < 3/2$ and $K < 2$). Let us now study
this problem using a variational method
\cite{mezard_variational_global}. We first use replicas to average over
disorder in (\ref{bosonized-hamiltonian}). We then approximate the
replicated action by the best quadratic action $S_0 = \frac 1 2
\int_{q,\omega_n} \phi^a_{q,\omega_n}
G_{ab}^{-1}(q,\omega_n)\phi^b_{q,\omega_n}$ where the $G$ are the
variational parameters and $a,b$ replica indices
\cite{giamarchi_columnar_variat}. Skipping the technical details which
can be found in \cite{orignac_commensurate_long}, we find that the
saddle point solution yields \widetext
\begin{equation} \label{eq:solution}
v G_c^{-1}(q,\omega_n) = \frac1{\pi  \overline{K}} ( \omega_n^2 + v^2
q^2) + m^2 + \Sigma_1(1-\delta_{n,0}) + I(\omega_n)
\end{equation}
\narrowtext
where $G_c = \sum_b G_{ab}$ is the connected Green function.
The parameters $m$, $\Sigma_1$ and the function $I(\omega_n)$ obey
closed self
consistent equations. The important physical quantities are simply
given in term of (\ref{eq:solution}): the static compressibility
reads
$\kappa = \lim_{q\to 0} \lim_{\omega \to 0} q^2 G_c(q,\omega)$
while the conductivity is given by the analytical continuation to
real frequencies:
$\sigma(\omega) = \left. [\omega_n \lim_{q\to 0}  G_c(q,\omega_n)]
\right|_{i\omega_n \to \omega + i \delta}$.
Note that both quantities are stemming from the
\emph{same} propagator but with  \emph{different}  limits
$q\to 0$ and $\omega \to 0$.

Since we expect the physics to be continuous for small enough $K$
(i.e. repulsive enough interactions), one can gain considerable
insight by considering the classical limit $\hbar \to 0$, $K\to 0$
keeping $\overline{K}=K/\hbar$ fixed. In this limit one can solve
analytically the saddle point equations and compute $m$,
$\Sigma_1$ and $I(\omega_n)$.
The resulting phase diagram is parameterized with
two physical lengths (for $K\to 0$): The  correlation length (or
soliton size) of the
pure gapped phase  $d=((4 g \overline{K})/(\alpha v))^{-1/2}$
\cite{giamarchi_mott_shortrev} and the
localization (or pinning) length $l_0=((\alpha v)^2/(16 W
\overline{K}^2))^{1/3}$ \cite{giamarchi_loc_ref,fukuyama_pinning} in the
absence of commensurability.
We find three phases as shown in Figure~\ref{fig:phases}:

\emph{Mott insulator} [MI]: At weak disorder we find a replica
symmetric solution with $\Sigma_1 = 0$ but with $m \ne 0$ ($m$ depends
on the disorder). $m \ne 0$ leads to zero compressibility $\kappa
= 0$. $m$ defines the correlation length $\xi$, with
$\xi^2 = v^2/(\pi \overline{K} m^2)$, in the
presence of both the disorder and the commensurate potential.
The effect of disorder is to increase $\xi$ compared to the pure
case, since it reduces the gap created by the commensurate
potential. We find that $\xi$ is given by
$(d/\xi)^2\exp[\frac14(\xi/l_0)^3]=1$. We obtain $\sigma(\omega)=0$ if $\omega < \omega_c= m\sqrt{1+\lambda - 3(\lambda/2)^{2/3}}$ where
$\lambda =(\xi/l_0)^3$.
The physics of this phase is similar to the simple
Mott insulator. However the gap in the conductivity decreases,
when disorder increases, and closes for $\lambda = 2$.
For $\lambda > 2$ ($d/l_0 > 0.98$) the RS solution becomes unphysical
\emph{even though} the mass $m$ remains finite at this transition point.

For stronger disorder one must break replica symmetry.
In the absence of commensurate
potential such a solution describes well the $d=1$ Anderson
insulator\cite{giamarchi_columnar_variat}, in which
$\Sigma_1 \ne 0$. Here, however \emph{two}
possibilities arise depending on whether
the saddle point allows for $m \ne 0$ or not:

\emph{Anderson Glass} [AG]: For large disorder compared to the
commensurate potential $d/l_0 >  1.58$, $m=0$ is the only saddle
point solution. In this case one recovers exactly the solution of
the Anderson insulator with interactions but no
commensurate potential. We call it Anderson glass
to emphasize that this phase is dominated by disorder
(the corresponding phase for bosons, also described by
(\ref{bosonized-hamiltonian}),
is the Bose glass \cite{giamarchi_loc_ref,fisher_bosons_scaling}).
Within the variational approach, the AG has a finite compressibility (identical to the one of the pure
system $\kappa = \pi\overline{K}/v$) and
the conductivity starts
as $\sigma(\omega) \sim \omega^2$ showing no gap.
Physically this is what is naively expected if the disorder washes
out completely the commensurate potential. While the MI and AG
were the only two phases accessible by previous techniques
\cite{mori_scba,fujimoto_mott+disorder_1ch} we find that an
intermediate phase exists between them.

\begin{figure}
 \centerline{\epsfig{file=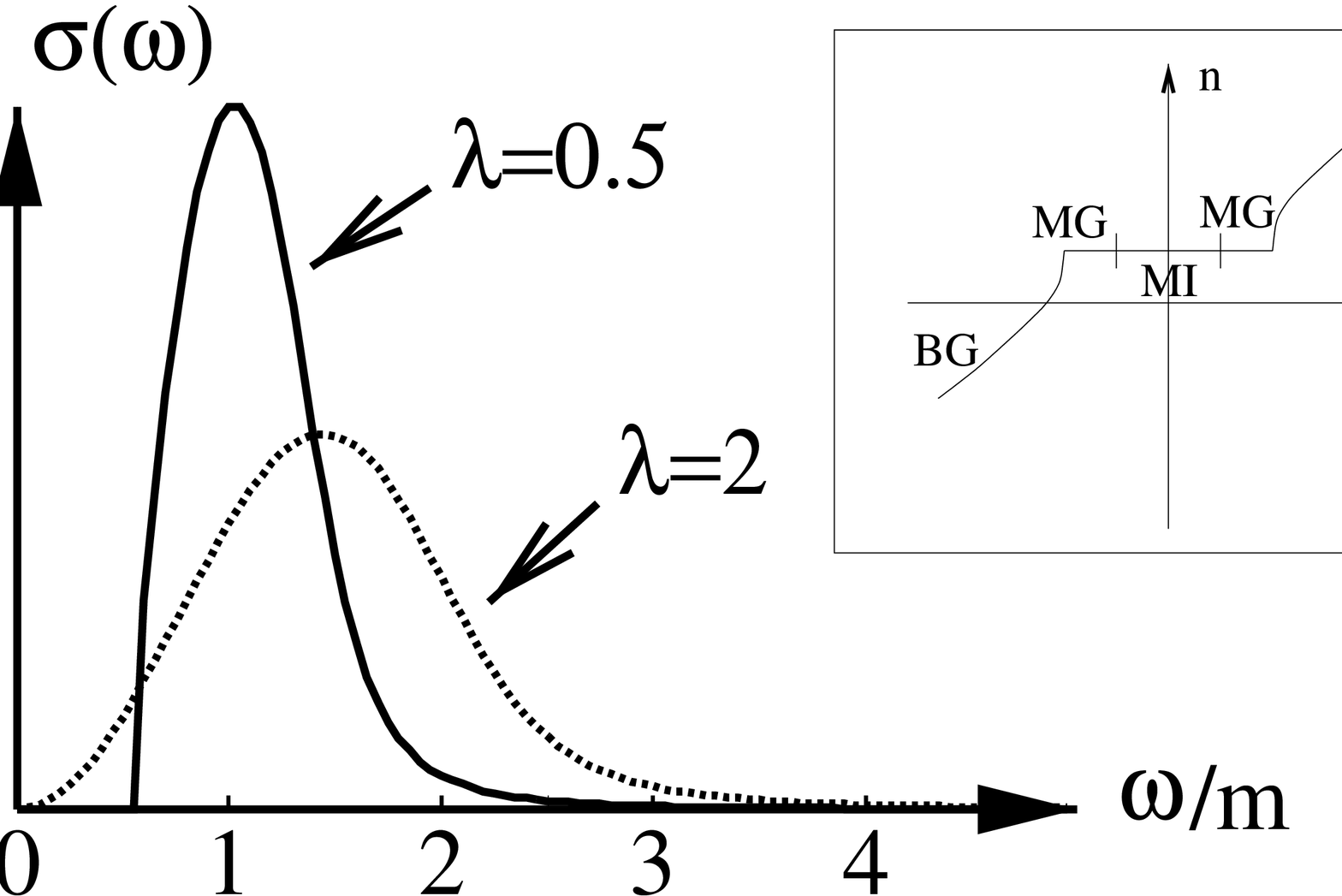,angle=0,width=7cm}}
  \caption{\label{fig:conduc} Conductivity in the MI (solid line)
   and MG phases (dashed line). Insert: density $n$ versus the chemical
  potential $\mu $.}
\end{figure}

\emph{Mott Glass} [MG]:
For intermediate disorder $0.98 < d/l_0 < 1.58$ a phase
with \emph{both} $\Sigma_1 \ne 0$ and  $m \ne 0$ exists. $m$ and
$\Sigma_1$ define two characteristic lengths in the intermediate
phase, $m^2 + \Sigma_1$ remaining constant in the MI and MG
\cite{orignac_commensurate_long}. On the other hand $m$ varies and
vanishes (discontinuously within the variational method)
at the transition from MG to AG.
The MG is thus neither a Mott nor an Anderson insulator. In
particular, the optical conductivity has \emph{no gap} for small
frequencies $\sigma(\omega) \sim \omega^2$, while
due to $m \ne 0$, the system is \emph{incompressible}
$\kappa = 0$. These properties are shown in Fig.~\ref{fig:conduc}.

This result is quite remarkable since by analogy with  noninteracting
electrons one is tempted to associate
a zero compressibility to the absence of available states at the
Fermi level and hence to a gap in the
conductivity as well. Our solution shows this is not the case,
when interactions
are turned on  the excitations
that consists in adding one particle (the important ones for the
compressibility) become  quite different from the particle hole
excitations that dominate the conductivity.

Physical arguments are also in favor of the existence of the Mott Glass,
both for systems with or without spins. Let us consider
the atomic limit, where the
hopping is zero. If the repulsion extends over at least one
interparticle distance, leading to small values of $K$,
particle hole excitations are
lowered in energy by excitonic effects.
For example for fermions with spins with both an onsite $U$ and
a nearest neighbor $V$ the gap to add one particle is
$\Delta = U/2$. On the other hand the minimal
particle-hole excitations  would be to have the
particle and hole on neighboring sites (excitons) and cost
$\Delta_{\text{p.h.}}=U - V$.
When disorder is added  the gaps  decrease
respectively \cite{footnote_bounded_disorder} as $\Delta \to
\Delta -W$ and $\Delta_{\text{p.h.}} \to \Delta_{\text{p.h.}} - 2W$.
Thus the conductivity gap  closes,
the compressibility remaining zero \cite{footnote_wigner_argument}.
According to this physical picture of the MG, the low frequency behavior of
conductivity is dominated by excitons (involving neighboring
sites). This is at variance from the AG where the particle and
the hole are created on distant sites. This may have consequences
on the precise low frequency form of the conductivity such as
logarithmic corrections.
When hopping is restored, we expect the excitons to dissociate
and the MG to disappear above a critical value $K>K^*$.
Since finite range is needed for the interactions,
in all cases (fermions or bosons) $K^* < 1$. In addition we expect
$K^* < 1/2$ for fermions with spins.
Similar excitonic arguments should also hold
in in two- or three-dimensional bosonic and fermionic Mott
insulators provided some finite range of interaction is taken into
account. In higher dimension, since disorder has a weaker impact  on
the transport properties, one expects that the important change
in the conductivity occurs at the transition between the MI
and MG, whereas the compressibility would become non zero only for stronger
disorder (transition MG to AG).  Numerical investigations would prove
valuable. Small gaps facilitate the observation of MG physics
(see Fig.~\ref{fig:conduc}), making the study of systems
already close to a metal -- Mott insulator transition
particularly interesting .

The properties that we have obtained are thus quite general,
depending only on the two gaps  closing separately. Given the
mapping between a $d$ dimensional quantum problem and a $d+1$ classical
one our study also applies to commensurate classical systems
in presence of disorder {\it correlated}
in at least one direction (here the imaginary time $\tau$).
(\ref{bosonized-hamiltonian}) can be generalized to any dimension
to describe a classical elastic system where $\phi$ becomes a displacement
field $u$. It applies to systems with internal periodicity, such as
a crystal or a charge density wave (with $2 \phi = K_0 u$ for
reciprocal lattice vector $K_0$)
or to non periodic systems such as interfaces. The periodic potential
 making the system flat while the disorder  makes it rough.
For these systems  a functional RG procedure (FRG) in $d=4-\epsilon$ (i.e
near $4+1$-dim systems) can be used.
For {\it uncorrelated} disorder in the absence of an external
periodic potential an internally periodic system
is described in $d \leq 4$ by a $T=0$ ``Bragg glass'' fixed point
\cite{giamarchi_vortex_long}.
Adding a periodic potential $\cos(p \phi)$ as a
perturbation,  a transition
was found at $T=0$ \cite{emig_commdisorder_frg} between the Bragg glass
(for $p>p_c(d)$)
(periodic potential irrelevant) and a commensurate phase (for $p
<p_c(d)$) (periodic potential relevant).
Such a $T=0$ transition also exists for correlated
disorder \cite{orignac_commensurate_long}.

To confirm the existence of the MG phase
it is necessary to study the phase
where the periodic potential is relevant (since $p=2<p_c(d)$ here).
Since this goes beyond the perturbative
FRG analysis, we consider the toy model where the
cosine is replaced by a quadratic term (a good approximation
when the periodic potential is relevant) defined by the energy:
\begin{eqnarray}
\frac{H}{T} = \frac{1}{T} \int d^d x d\tau [\frac{1}{2} ( c (\nabla u)^2
+ c_{44} (\partial_\tau u)^2 + m^2 u^2 ) + V(x,u(x,\tau)) ]
\end{eqnarray}
where $\tau$ is the coordinate along which disorder is correlated
(e.g. the magnetic field for vortices with columnar defects), $T$ is the
classical temperature ($\hbar$ for the quantum problem)
and the gaussian disorder has a correlator
$\overline{V(x,u) V(x',u')}= \delta^d(x-x') R(u-u')$ (for
a periodic problem, $R(u)$ is itself periodic).
$c$ and $c_{44}$ are the elastic moduli, analogous to
$1/\overline{K}$ for the quantum problem (\ref{bosonized-hamiltonian}).
This $d+1$ dimensional model can be studied perturbatively (for small
$m$ and disorder).
For $m=0$ and $T=0$, both for uncorrelated \cite{fisher_functional_rg} and
correlated disorder \cite{balents_loc}
a cusp-like {\it nonanalyticity} in the renormalized disorder $R(u)$
develops beyond the Larkin length $R_c$ (corresponding to
metastability and barriers in the dynamics) \cite{footnoteomega0}.
Remarkably, we find that
this feature \emph{persists} even when $m>0$,
while one usually expects that a mass smoothes out singularities.
This can be seen from the RG equation for
$\Delta(u)=-R''(u)$:
\widetext
\begin{eqnarray}
\partial_l \Delta(u) &=& \epsilon \Delta(u) + \tilde{T}_l \Delta''(u) +
f_l (\Delta''(u) (\Delta(0) - \Delta(u)) - \Delta'(u)^2)
\label{frg}
\end{eqnarray}
\narrowtext with $f_l=\frac1{8\pi^2}(1 + \mu  e^{2 l})^{-2}$, $\mu =m^2
a^2$ and at zero temperature $T_l=0$. Integrating the closed equation
for $\Delta''(0)$ one finds that the cusp persists (i.e
$-\Delta''_{l=+\infty}(0) = +\infty$) provided that $R_c < R_c^*(\mu )$
(with $R_c^*(\mu  ) \sim 1/\sqrt\mu    $ for small $\mu    $) while it
is washed out ($-\Delta''_{l=+\infty}(0) < + \infty$) for weaker
disorder.  Our FRG study shows that this $T=0$ transition in the
renormalized disorder, exists \emph{both} for correlated and
uncorrelated disorder. For uncorrelated disorder our findings are of
interest for the question of the existence of an intermediate ``glassy
flat phase''. In that case, however, no sharp signature of this
transition exists in two point correlation functions and a physical
order parameter remains to be found, which makes the existence of such
a phase still controversial
\cite{bouchaud_commdisorder_variational,emig_commdisorder_long}. On the
contrary for correlated disorder the transition seen in the FRG has
much stronger physical consequences. Because of the lack of rotational
invariance (in $(x,\tau)$) the existence of the cusp and the transition
directly affects two point correlation functions. Indeed, the tilt
modulus $c_{44}$ renormalizes as $\partial_l c_{44} = - f(l)
\Delta''(0) c_{44}$. Integrating at $T=0$ one finds that
$c_{44}(l=+\infty)$ is finite for $R_c > R_c^*(\mu  )$ but that it is
infinite for $R_c < R_c^*(\mu  )$. Furthermore we find (see
\cite{orignac_commensurate_long} for details) that for correlated
disorder a small temperature ($\tilde{T}_l >0$) does not affect the
transition (the phase where $c_{44}(l=+\infty)=\infty$ survives)
whereas for uncorrelated disorder the effective temperature goes to a
constant ($\sim \mu  ^{1-\epsilon/2}$) washing out the cusp. Thus the
toy model for correlated disorder exhibits at low temperature, within
the FRG, a transition between two phases. The first one is identified
with the MI, where the mass (commensurability) destroys the
metastability and restores isotropy in $x,\tau$
 at large scale. The second one is the
MG, which is glassy with metastable states
despite the presence of the mass. It shares some properties
with the AG such as $c_{44} = \infty$. This implies non
analyticity of the Green's functions in frequency.
The AG itself corresponds
to the phase where the periodic potential is irrelevant ($m=0$).
This provides strong evidence for the  intermediate phase
proposed in this paper, which, besides electrons, should be obtained
in classical commensurate elastic systems with disorder.

\begin{thebibliography}{10}
\bibitem{periodic_mott}
In one dimension, the Mott transition can be viewed as caused by a $4k_F$
  periodic potential (see e.g. \cite{giamarchi_mott_shortrev}). However to
  study the simpler case of spinless fermions (while retaining the main
  features of the fermions with spins), we use instead a $2 k_F$ periodic
  potential which allows to study gap formation while avoiding the destruction
  of the gap by infinitesimal disorder that occurs in the latter case
  \cite{shankar_spinless_conductivite}.
\bibitem{footnote_bounded_disorder}
For small bounded disorder the gap of the Mott insulator phase is robust.
\bibitem{footnoteomega0}
Note that at $T=0$ the $\omega_n=0$ mode (i.e $u(x,\tau)$ constant in $\tau$)
  decouples and identifies exactly with the problem of uncorrelated disorder in
  $d$ dimension.
\end{thebibliography}
\end{document}